\documentclass[aps,preprint,onecolumn,showpacs]{revtex4}

\usepackage{graphicx}
\usepackage{dcolumn}
\usepackage{bm}

\begin{document}

\title{Reshape the perfect electrical conductor cylinder at will}

\author{Huanyang Chen$^*$, Xiaohe Zhang, Xudong Luo, and
Hongru Ma}

\affiliation{Department of Physics, Shanghai Jiao Tong University,
Shanghai 200240, China}

\author{C.T. Chan}

\affiliation{Department of Physics, The Hong Kong University of
Science and Technology, Clear Water Bay, Hong Kong, China}

\begin{abstract}
A general method is proposed to design the cylindrical cloak,
concentrator and superscatterer with arbitrary cross section. The
method is demonstrated by the design of a perfect electrical
conductor (PEC) reshaper which is able to reshape a PEC cylinder
arbitrarily by combining the concept of cloak, concentrator and
superscatterer together. Numerical simulations are performed to
demonstrate its properties.
\end{abstract} \maketitle


The pioneer work on transformation optics and cloaking [1-4] has
suggested a new kind of methodology to manipulate the
electromagnetic (EM) waves by means of the metamaterials. Based on
the transformation optics, various exciting functional devices have
been designed both in theory [5-15] and in experiment [16]. Chen
\textit{et al.} has recently proposed the ``imperfect cloak'' [7]
which can reduce the scattering cross section of an object. Start
from this imperfect cloak, they also suggested a design of
``dispersive cloak''. In contrast to the imperfect cloak, a concept
of ``superscatterer'' [12] was also proposed which can obtain a
giant scattering cross section beyond the device itself. If the
inner core of the concentrator [11] is replaced with a PEC cylinder,
the concentrator can employed as a device to change the scattering
cross section of the PEC cylinder. And its functionality is between
the imperfect cloak and the superscatterer. In addition, a
cylindrical cloak of arbitrary cross section [10] has been given,
which can be easily extended to the imperfect cloak, concentrator,
and superscatterer. In this paper, we will show the method of such
an extension. And based on our methodology, we will propose a kind
of transformation media, which we shall call the ``PEC reshaper''.
An explicit design will be given. The PEC reshaper can reshape the
PEC cylinder at will. For example, the effective PEC cylinder can be
partially inside the device and partially outside the device. We
will demonstrate the properties of the PEC reshaper by using the
finite-element methods.

\begin{figure}
\begin{center}
\includegraphics[angle=-0,width=1.0\columnwidth] {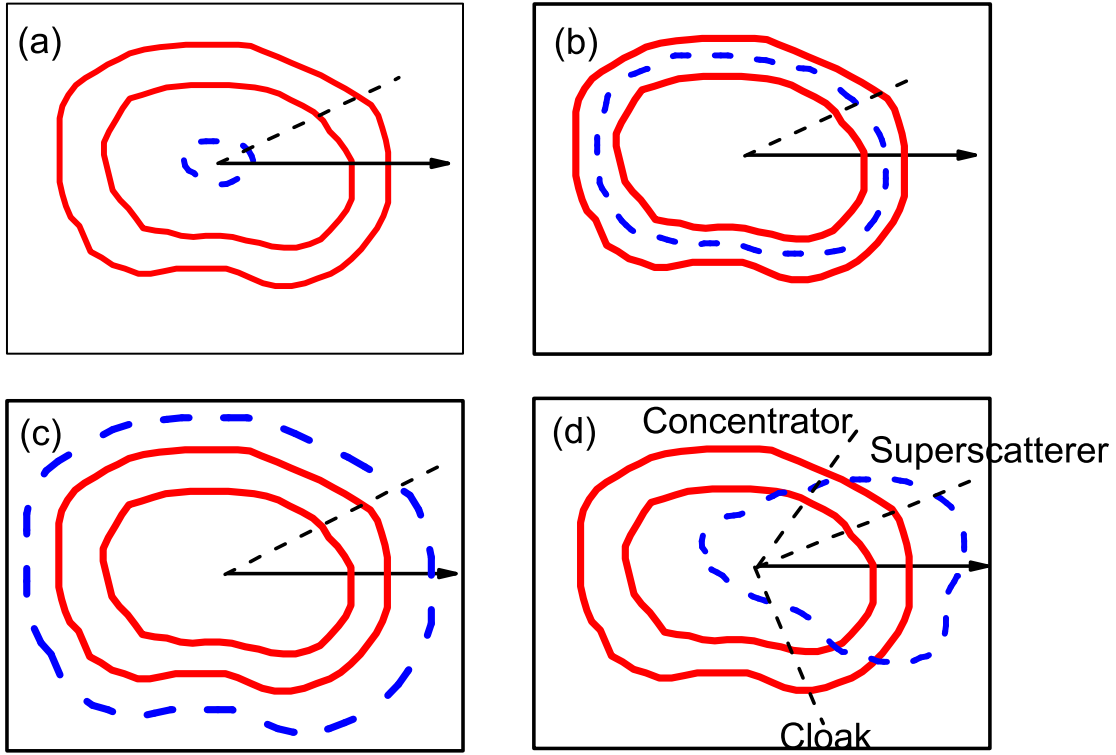}
\end{center}
\caption{(Color online) The schematic plot for the coordinate
transformation of (a) an imperfect cylindrical cloak with arbitrary
cross section, (b) a cylindrical concentrator with arbitrary cross
section, (c) a cylindrical superscatterer with arbitrary cross
section, and (d) a cylindrical PEC reshaper. The red solid lines
denote the outer and inner boundaries of the above devices, while
the blue dashed lines outline the effective PEC boundaries in the
view of the outside world.}\label{fig.1}
\end{figure}

The general coordinate transformation is given by the following
relation,

\begin{equation}
\label{eq1} \left\{ {{\begin{array}{*{20}c}
 {r'(r,\theta ) = \frac{b - a}{b - c}(r - b) + b = \frac{b - a}{b - c}r +
\frac{a - c}{b - c}b,} \hfill \\
 {\theta ' = \theta ,\mbox{ }0 \le \theta < 2\pi } \hfill \\
 {z' = z,\mbox{ }z \in {\rm R}} \hfill \\
\end{array} }} \right.,
\end{equation}

\noindent where $a = \rho _1 (\theta ),b = \rho _2 (\theta ),c =
\rho _3 (\theta ),$ which are three functions that specify the inner
and outer boundaries of the transformation media and the boundary of
an imaginary cylinder which will be useful later, respectively. We
note that all the boundaries are smooth and non-convex. This
transformation maps the field in the domain $\rho _3 (\theta ) < r <
\rho _2 (\theta )$ onto another domain $\rho _1 (\theta ) < r' <
\rho _2 (\theta )$. From the transformation optics, one can obtain
the parameters inside the transformation media in $\rho _1 (\theta )
< r' < \rho _2 (\theta )$. We consider the transverse electric (TE)
mode for simplicity and suppose that the inner cylinder ($r' \le
\rho _1 (\theta ))$ is a PEC cylinder. In the view of transformation
optics, after coated with the transformation media in $\rho _1
(\theta ) < r' < \rho _2 (\theta )$, the inner PEC cylinder ($r' \le
\rho _1 (\theta ))$ looks like another effective PEC cylinder in the
domain $r \le \rho _3 (\theta )$. If $\rho _3 (\theta ) \le \rho _1
(\theta )$, the transformation media is an imperfect cylindrical
cloak [7, 8] with arbitrary cross section, which can reduce the
cross section of the inner PEC cylinder (See in Fig. 1(a)). If $\rho
_3 (\theta ) \equiv 0$, the imperfect cylindrical cloak becomes
perfect [1, 10]. If $\rho _1 (\theta ) \le \rho _3 (\theta ) \le
\rho _2 (\theta )$, the transformation media is the outer shell of a
cylindrical concentrator [11] with arbitrary cross section, which
can enhance the cross section of the inner PEC cylinder (See in Fig.
1(b)). We shall still call the device ``concentrator'' for
simplicity. However, the effective PEC cylinder is still within the
domain $r' \le \rho _2 (\theta )$. With the help of the concept of
``superscatterer'' [12], we can reshape the inner PEC cylinder very
versatilely. In the case of superscatterer (i.e. $\rho _2 (\theta )
\le \rho _3 (\theta ))$, the transformation maps the field in the
domain $\rho _2 (\theta ) < r < \rho _3 (\theta )$ onto another
domain $\rho _1 (\theta ) < r' < \rho _2 (\theta )$ through the
mirror of $r' = r = \rho _2 (\theta )$. The superscatterer can
enhance the cross section of the inner PEC cylinder to that of an
effective PEC cylinder which can be much larger than the device (See
in Fig. 1(c)). Now we consider an extraordinary case shown in Fig.
1(d), where the effective PEC cylinder is partially inside the inner
cylinder, partially within the domain of transformation media and
partially outside the device for different angles. Such a device
combines cloak, concentrator and superscatterer all together into
one case, which we shall call the ``PEC reshaper'' as it could
reshape the PEC boundary more or less arbitrarily.

Before demonstrating the properties of the PEC reshaper, we have to
work out the required parameters for it from the transformation
optics. The Jacobian is

\begin{equation}
\label{eq2} J = \left[ {{\begin{array}{*{20}c}
 {\frac{\partial r'}{\partial r}} \hfill & {\frac{\partial r'}{r\partial
\theta }} \hfill & 0 \hfill \\
 {\frac{r'\partial \theta '}{\partial r}} \hfill & {\frac{r'\partial \theta
'}{r\partial \theta }} \hfill & 0 \hfill \\
 0 \hfill & 0 \hfill & 1 \hfill \\
\end{array} }} \right],
\end{equation}

\noindent where $\frac{\partial r'}{\partial r} = \frac{b - a}{b -
c}$, $\frac{r'\partial \theta '}{\partial r} = 0$, $\frac{r'\partial
\theta '}{r\partial \theta } = \frac{r'}{r}$, and

\begin{equation}
\label{eq3} \frac{\partial r'}{r\partial \theta } = \frac{a'(c - b)
+ b'(a - c) + c'(b - a)}{r(b - c)}\frac{r' - b}{b - a} + \frac{a -
c}{b - c}\frac{b'}{r},
\end{equation}

\noindent with $\rho ' = \frac{\partial \rho }{\partial \theta
},\rho = a,b,c, \quad r = \frac{b - c}{b - a}r' + \frac{c - a}{b -
a}b$, and $\theta = \theta '$.

The corresponding permittivity and permeability tensors of the
transformation media on the circular cylindrical coordinate are
[13],

\begin{equation}
\label{eq4}
\mathord{\buildrel{\lower3pt\hbox{$\scriptscriptstyle\leftrightarrow$}}\over
{\varepsilon }} _c =
\mathord{\buildrel{\lower3pt\hbox{$\scriptscriptstyle\leftrightarrow$}}\over
{\mu }} _c = JJ^T / \det (J) = \left[ {{\begin{array}{*{20}c}
 {\mu _{rr} } \hfill & {\mu _{r\theta } } \hfill & 0 \hfill \\
 {\mu _{\theta r} } \hfill & {\mu _{\theta \theta } } \hfill & 0 \hfill \\
 0 \hfill & 0 \hfill & {\varepsilon _{zz} } \hfill \\
\end{array} }} \right],
\end{equation}

\noindent where $\mu _{rr} = \frac{(\frac{\partial r'}{\partial
r})^2 + (\frac{\partial r'}{r\partial \theta })^2}{\frac{\partial
r'}{\partial r}\frac{r'}{r}}, \quad \mu _{r\theta } = \mu _{\theta
r} = \frac{\frac{\partial r'}{r\partial \theta }}{\frac{\partial
r'}{\partial r}}, \quad \mu _{\theta \theta } =
\frac{\frac{r'}{r}}{\frac{\partial r'}{\partial r}},$ $\varepsilon
_{zz} = \frac{1}{\frac{\partial r'}{\partial r}\frac{r'}{r}}.$ We
note that the permittivity and permeability tensors here should be
written as the functions of positions in the form of $(r',\theta
')$.

Consider the TE mode, we have, $\varepsilon _{zz} =
\frac{1}{\frac{\partial r'}{\partial r}\frac{r'}{r}},$ and

\begin{equation}
\label{eq5}
\begin{array}{l}
 \left[ {{\begin{array}{*{20}c}
 {\mu _{xx} } \hfill & {\mu _{xy} } \hfill \\
 {\mu _{xy} } \hfill & {\mu _{yy} } \hfill \\
\end{array} }} \right] = \left[ {{\begin{array}{*{20}c}
 {\cos \theta '} \hfill & { - \sin \theta '} \hfill \\
 {\sin \theta '} \hfill & {\cos \theta '} \hfill \\
\end{array} }} \right]\left[ {{\begin{array}{*{20}c}
 {\mu _{rr} } \hfill & {\mu _{r\theta } } \hfill \\
 {\mu _{r\theta } } \hfill & {\mu _{\theta \theta } } \hfill \\
\end{array} }} \right]\left[ {{\begin{array}{*{20}c}
 {\cos \theta '} \hfill & {\sin \theta '} \hfill \\
 { - \sin \theta '} \hfill & {\cos \theta '} \hfill \\
\end{array} }} \right] \\
 = \left[ {{\begin{array}{*{20}c}
 {\mu _{rr} \cos ^2\theta ' - 2\mu _{r\theta } \sin \theta '\cos \theta ' +
\mu _{\theta \theta } \sin ^2\theta '} \hfill & {(\mu _{rr} - \mu
_{\theta \theta } )\sin \theta '\cos \theta ' + \mu _{r\theta }
(\cos ^2\theta ' -
\sin ^2\theta ')} \hfill \\
 {(\mu _{rr} - \mu _{\theta \theta } )\sin \theta '\cos \theta ' + \mu
_{r\theta } (\cos ^2\theta ' - \sin ^2\theta ')} \hfill & {\mu _{rr}
\sin ^2\theta ' + 2\mu _{r\theta } \sin \theta '\cos \theta ' + \mu
_{\theta
\theta } \cos ^2\theta '} \hfill \\
\end{array} }} \right], \\
 \end{array}
\end{equation}

\noindent on Cartesian coordinate [14]. We note that the above
parameters are also valid for cloak, concentrator and superscatterer
with arbitrary cross section.

\begin{figure}
\begin{center}
\includegraphics[angle=-0,width=0.45\columnwidth] {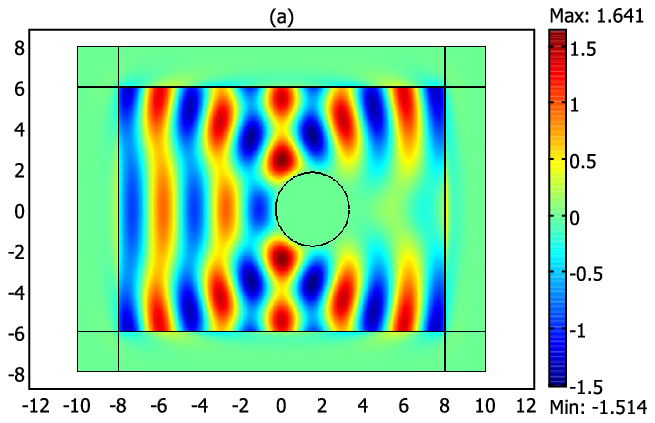}
\includegraphics[angle=-0,width=0.45\columnwidth] {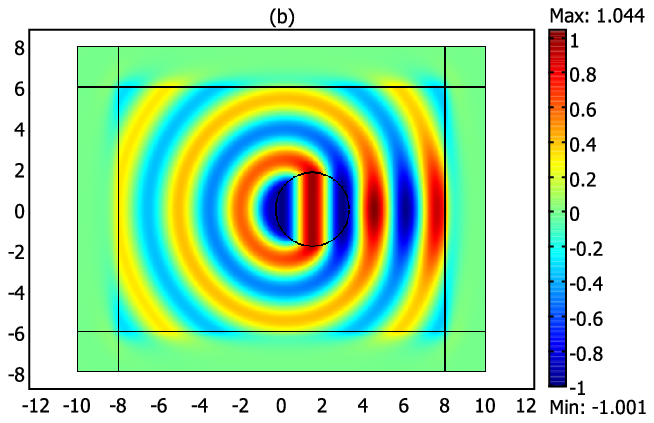}\\
\includegraphics[angle=-0,width=0.45\columnwidth] {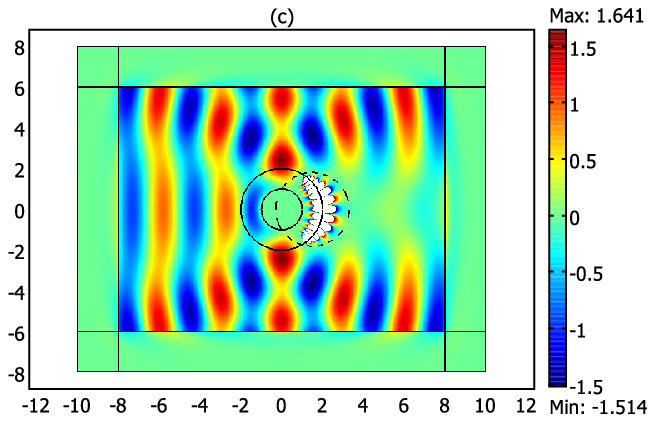}
\includegraphics[angle=-0,width=0.45\columnwidth] {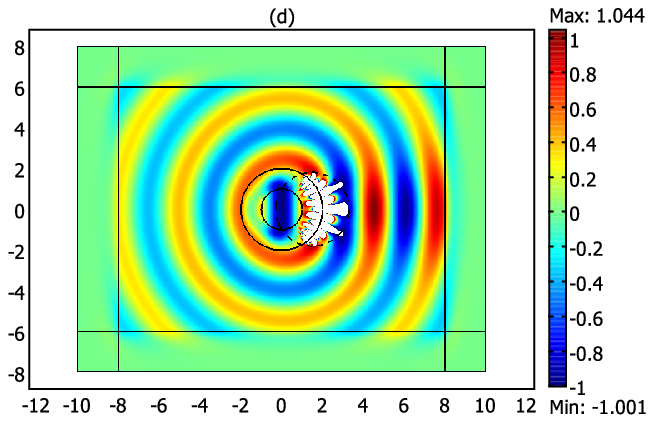}
\end{center}
\caption{(Color online) Snapshot of the total and scattering
electric field. (a)-(b) The total and scattering electric field
caused by a PEC cylinder with its outer boundary depicted by $c =
\rho _3 (\theta )$, respectively. (c)-(d) The total and scattering
electric field induced by the designed PEC reshaper, respectively.
The dashed lines in (c) and (d) outline the boundary of the
effective PEC cylinder, which is the same to the outer boundary of
the PEC cylinder in (a) and (b).}\label{fig.2}
\end{figure}

With the explicit form of the required parameters, the properties of
the PEC reshaper can be demonstrated by numerical simulations with
the COMSOL Multiphysics finite element-based electromagnetics
solver. Let $a = \rho _1 (\theta ) = 1m,b = \rho _2 (\theta ) = 2m$,
and $c = \rho _3 (\theta ) = x_0 \cos \theta + \sqrt {a^2 + x_0^2
\cos ^2\theta } ,$ where $x_0 = \frac{b^2 - a^2}{2a}$. We choose $c
= \rho _3 (\theta )$ as a circle for simplicity. A plane wave is
normal incident from left to right with unit amplitude and a
frequency of 0.1 GHz. Fig. 2(a) and (b) show the snapshots of the
total electric and scattering field caused by a PEC cylinder with
its outer boundary depicted by $c = \rho _3 (\theta )$,
respectively. Fig. 2(c) and (d) show the total electric and
scattering field induced by the PEC reshaper which is a concentric
cylindrical shell. Comparing the similar far-field pattern from Fig.
2(a) and (c), or Fig. 2(b) and (d), we can conclude that the PEC
reshaper reshapes the PEC cylinder with $a = \rho _1 (\theta ) = 1m$
to an effective PEC cylinder with $c = \rho _3 (\theta )$. The large
overvalued fields in Fig. 2(c) and (d) are caused by the surface
mode resonances excited at $c = \rho _3 (\theta )$, which are
replaced with white flecks [12].

In conclusion, we have proposed the concept of PEC reshaper which
combines the concept of cloak, concentrator and superscatterer all
together into one case. The PEC reshaper can reshape the PEC
boundary arbitrarily if the objective effective PEC cylinder shares
a domain with the original PEC cylinder (i.e. there exists $\theta '
= \theta \in [0,\;2\pi ))$. The properties of the PEC reshaper are
demonstrated by numerical simulations from finite-element methods.
The design can be extended to three dimensions straitforwardly.

\section*{Acknowledgments}
This work was supported by the National Natural Science Foundation
of China under grand No.10334020 and in part by the National
Minister of Education Program for Changjiang Scholars and Innovative
Research Team in University, and Hong Kong Central Allocation Fund
HKUST3/06C.\\
$^*$Correspondence should be addressed to: kenyon@ust.hk

\end{document}